# Designing Chatbots to Support Victims and Survivors of Domestic Abuse

Rahime Belen Saglam[a1], Jason R. C. Nurse[b], Lisa Sugiura[c]

[a] *School of Architecture, Computing and Engineering, University of East London, UK*
[b] *School of Computing, University of Kent, UK*
[c] *School of Criminology and Criminal Justice, University of Portsmouth, UK*
[1] rbelen@uel.ac.uk

**Objective:** Domestic abuse cases have risen significantly over the last four years, in part due to the COVID-19 pandemic and the challenges for victims and survivors in accessing support. In this study, we investigate the role that chatbots – Artificial Intelligence (AI) and rule-based – may play in supporting victims/survivors in situations such as these or where direct access to help is limited.
**Methods:** Interviews were conducted with experts working in domestic abuse support services and organizations (e.g., charities, law enforcement) and the content of websites of related support-service providers was collected. Thematic content analysis was then applied to assess and extract insights from the interview data and the content on victim-support websites. We also reviewed pertinent chatbot literature to reflected on studies that may inform design principles and interaction patterns for agents used to support victims/survivors.
**Results:** From our analysis, we outlined a set of design considerations/practices for chatbots that consider potential use cases and target groups, dialog structure, personality traits that might be useful for chatbots to possess, and finally, safety and privacy issues that should be addressed. Of particular note are situations where AI systems (e.g., ChatGPT, CoPilot, Gemini) are not recommended for use, the value of conveying emotional support, the importance of transparency, and the need for a safe and confidential space.
**Conclusion:** It is our hope that these considerations/practices will stimulate debate among chatbot and AI developers and service providers and – for situations where chatbots are deemed appropriate for use – inspire efficient use of chatbots in the support of survivors of domestic abuse.

## 1. Introduction

Domestic abuse is an appalling crime. Globally, 35 per cent of women have ever experienced physical or sexual abuse.[38] In the year ending March 2020 in England and Wales, 2.4 million adults aged 16-74 years experienced domestic abuse (1.6 million women, 786,000 men).[11] It is estimated that worldwide less than 40 per cent of the women who experience violence seek help of any sort.[38] However, online interventions are limited in some regards, for instance, women who had left an abusive situation can access to very few guidance compared to the ones who are leaving or preparing to leave an abusive relationship.[29] On the other hand, many cases remain hidden since disclosure is difficult and may result in additional trauma due to inadequate support.[6] In addition to victim's concerns about social stigmatization and the potential for perpetrator retribution, characteristics of the environment are also reported to create barriers during the disclosure process. Formal proceedings, inappropriate responses to disclosure including victim blaming, may cause further discomfort and distress.[6]

Today, technology is playing an increasingly important role in supporting victims of domestic abuse. This is especially the case under the conditions of COVID-19 and various national lockdowns, where technology solutions that can scale and allow survivors to be supported in their own environments are vital. For instance, websites of supporting organizations provide expert opinions and advice to guide the survivors. Some supporting organizations also provide live chat options with their support workers on their websites, which solely depend on manual human effort. While there are not many, there are a few chatbot solutions that have been developed to provide essential information necessary for survivors as an alternative to live chats. Examples of these include the one provided by Refuge[28] which is dedicated to help safe use of smartphones, and the agent launched on the website of Southall Black Sisters[34] to provide support on domestic abuse. Refuge, a United Kingdom charity providing specialist support for women and children experiencing domestic abuse, provided a rule based chatbot on their websites to help victims of online abuse.[28] Another chatbot application, HelloCass[14], has been developed in Australia to provide information on domestic and family violence, sexual harassment, and sexual assault. A similar application was developed in Thailand, for survivors of sexual abuse that responds with information about how to report to the police, how to preserve evidence, and what support services or compensation they are entitled to by law.[37] Guidance on how to report incidents has been provided by another chatbot developers in India providing information on recording incidents of personal abuse and sexual harassment anonymously.[8]

Chatbots are computer programs that can simulate a conversation with a human user. They typically use user input to generate appropriate conversational responses or execute defined tasks. Today, chatbots are used in various industries including finance and health care. Their popularity has also increased significantly increased over the last year given the release of OpenAI's ChatGPT, Microsoft's CoPilot and Google's Gemini. Focusing on the literature, perhaps the mostly related chatbot solutions that may support survivors of domestic abuse are the ones developed for mental health. However, as the scoping study reveals, there is no study that specifically investigates and satisfies the needs of survivors of domestic abuse in the mental health domain[1].

Although existing applications summarized above are mainly rule based (scripted) and button based chatbots, there are more sophisticated Artificial Intelligence (AI) based chatbots proposed in the literature to support survivors of domestic or sexual abuse. In one of those studies, Bauer et al. implemented a chatbot to assist survivors of sexual harassment by directing them to appropriate institutions and to encourage them to report the incidents.[3] Applying classification algorithms on the responses of the users, harassment type has been identified (verbal abuse, non-verbal abuse, physical abuse, serious physical abuse, other) and location and date information extraction has been performed using Named Entity Recognition algorithms. Once all of the details were identified by the chatbot, depending on the type of abuse (physical, verbal, non verbal), the agent would provide specific information to the user depending on the case. Authors reported a success rate of more than 98% for the identification of a harassment-or-not case and around 80% for the specific type harassment identification. It was reported to achieve more than 90% accuracy for location and time. Furthermore, with the advent of Generative AI, including AI chatbots such as ChatGPT, CoPilot and Gemini, it is likely that these will be trialed for use in situations such as these, if only for immediate response before being passed to a victim support officer.

Park and Lee focused on disclosure of sexual violence and investigated if chatbots can lower the burden of this process.[23] The authors conducted interviews with eight police officers and two counselors and discovered the main burdens that victims frequently express while/before they report their case or get counseling are the significant amount of time required to report the case, risks of financial loss, emotional burdens (e.g., shame, guilt, anxiety, and fear), mental burdens (significant amount of attention and concentration needed) and privacy burdens. One of their interviewees stated that the necessity of a victims' personal information as a prerequisite to free counseling led to people leaving (without counseling) because they did not want to put it on record. Social burdens (risks of rumors, social loss, and social change) and physical burdens making victims physically uncomfortable are the other burdens reported by the researchers. Based on those findings, three design guidelines have been reported for chatbot developers: develop empathetic, and active listener applications; provide accurate information with speed, and finally ask personal information as minimally as possible.

While these articles are generally insightful, they offer little detailed information on design guidelines that can be used to inform new solutions. They are general meta-requirements which require further refinement to identify design principles that chatbot developers can easily understand and apply. Defining what are desirable characteristics of a chatbot in this context, the use cases that may be most appropriate; and outlining how agent interactions may be better designed to accommodate stakeholder needs still demands contextual investigations.

In this study, we investigate the extent to which design guidelines can be created for chatbots that could be used to support victims of domestic abuse. For this purpose, we conducted three related studies. Firstly, websites of UK-based organizations that support victims of domestic abuse and their features have been analyzed to determine the current status of the technology usage in this area. This enables us to understand organization's use cases, the values they promote, the people they target, how they are designed and how they interact. Here, we view websites as reasonable sources of input on "good practice" or things that need to exist. Secondly, literature has been reviewed to explore scientific approaches and technology-enabled solutions proposed to support victims of domestic abuse. In a parallel study, interviews with professionals working in UK-based supporting organizations have been conducted to gather data pertinent to our goal. Finally, we identified several design principles that could help shape several aspects including conversation design, user interfaces and personification of the chatbot.

This paper contributes to the existing state-of-the-art in chatbot research by guiding designers and developers on building appropriate solutions for victims of domestic abuse. Specifically, we provide insights into what use cases may be most appropriate, and how these agent interfaces and interactions can be better designed. Website analysis results, literature review outcomes and interview findings are mapped to guidelines for the design of chatbot interactions to better support victims/survivors and refined for preliminary design principles. Those principles can support future work by researchers and developers based at the intersections of domestic or online abuse and chatbots, as well as professionals working in domestic abuse services and those tackling it.

## 2. Study Design

This research study consists of three sub-studies. Firstly, current websites and the chatbots launched on those websites to provide support to abuse victims/survivors are identified; from this list, we then assess each one to determine the interfaces, tools and methods used to support victims/survivors of domestic abuse. To complement this work, we have reviewed literature to identify proposed guidelines and principles by researchers, given that academia is often more advanced in its use of high-tech solutions. As a third study, semi-structured interviews with experts working in support organizations have been conducted to gather qualitative, open-ended data pertinent to our goal. We note here that our study engages particularly with UK-based websites and UK-based interviewees and organizations. This is a point we reflect on later in the article, while also discussing the implications of our research. The study's data was collected in November 2020.

*2.1 Analysis of websites*

To analyze websites of organizations that support survivors of domestic abuse, we first needed to decide what organizations to choose. Due to the variety of organizations focusing on different target groups, there was no well-maintained list of such organizations with the needed indicators for us to consider. We therefore decided to use the list provided by the UK Citizens Advice[1] organization for victims of domestic abuse or violence. Citizens Advice is a network of independent charities throughout the UK that give free, confidential information and advice to assist people with financial, legal, consumer and other problems.[2] This list resulted in 19 organizations under five categories: organizations for women, organizations for men, organizations for women and men, organizations for lesbian, gay, bisexual and transgender people and finally, organizations for disabled people. We enriched this list with the websites of organizations that were suggested interviewees. The list of websites can be seen in Table 1.

Table 1: List of organizations

| Category | Name Organization |
| --- | --- |
| Organizations for women | National Domestic Abuse helpline/Refuge[28], Women's Aid[41], Rights of Women[21], Finding Legal Options for Women Survivors (FLOWS)[10], Southall Black Sisters[34], Respect - Men's Advice Line[19], ManKind Initiative[18], SurvivorsUK[35], Everyman Project[26] |
| Organizations for women and men | RCJ Advice Family Service[32], Rape Crisis[7], Honour Network Helpline[15], Action on Elder Abuse[22], National Stalking Helpline[17], Respect Phoneline[24] |
| Organizations for lesbian, gay, bisexual and transgender people | National LGBT+ Domestic Abuse Helpline[16] |
| Organizations for disabled people | SignHealth - Domestic Abuse Service[33], Respond[31] |

Each website has been analyzed to page depth of three; here, page depth refers to the number of clicks needed to reach a specific page from the homepage using the shortest path. The texts from pages were drawn together and the ones covering domestic abuse have been identified. Then, the collected contents were subjected to a thematic analysis. Braun and Clarke's guidelines were used for the analysis.[5]

In addition to the thematic analysis conducted on the content provided by websites, we have also observed the services provided to survivors by the organizations and the website design features adopted for those services. Their approaches to communicate with survivors, how they secure confidentiality of the data they collect (via live chats, forums etc.), existence of quick exit buttons or warnings that are given via pop-ups, have been noted where applicable.

---

[1] https://www.citizensadvice.org.uk/family/gender-violence/domestic-violence-and-abuse-getting-help/

*2.2 Interviews*

The goal of the interviews was to capture thoughts and perspectives of five experienced staff working in four supporting organizations. They had minimum of three, and a maximum of 14 years of experience in this particular domain. We have also conducted interviews with two police officers, with more than 10 years of experience. The research has been reviewed and ethically approved by the Research Ethics & Governance department of the University of Kent.

Interviews were audio recorded and transcribed by the authors. Initially, analysis was conducted to understand the data and reflect on its meaning. Secondly, the entire data set has been organized and analyzed following the same approach used for the website analysis.

*2.3 Analysis of the design guideline literature*

A search of chatbot-related papers on domestic abuse was conducted to construct a corpus of papers. The review included papers from IEEE Xplore, ACM Digital Library, SpringerLink and ScienceDirect retrieved using specific keywords given below: ("domestic abuse" OR "domestic violence" OR "coercive control" OR "intimate partner") AND ("chat bot" OR "chatbot" OR "conversational agent").

Titles and abstracts were examined to determine whether they discussed design guidelines or principles; research that fit these criteria were added to the corpus. Papers that reiterated existing research (e.g., previously published guidelines) or the ones that did not identify design guideline have been excluded at the further step where the full texts were examined. This led us to identify seven relevant papers out of 963. The distribution of the papers among the databases can be seen in Table 2.

Table 2: Paper analysis and filtering

| Search Engine / Digital Library | Number of Papers | Included |
|---|---|---|
| Springer | 244 | 2 |
| ScienceDirect | 106 | 0 |
| IEEE | 11 | 0 |
| ACM Digital Library | 9 | 1 |
| WoS | 1 | 0 |
| ProQuest | 118 | 1 |
| Google Scholar | 474 | 3 |
| **TOTAL** | **963** | **7** |

For each paper included in the corpus, we identified the main characteristics, design guidelines and principles proposed by the authors. We categorized our findings into three main groups: user-chatbot interaction, chatbot development and user experience as done in.[9] In the first category, suggested features for the bot responses were observed; chatbot personality, conversational flow, conversation length or dialog structure were noted. In the user-bot interaction category, principles regarding graphical appearance, task and duty specification, and possible interaction styles (button based, text based, voice based etc.) have been identified. Bot development category covered the design guidelines which identifies development details such as fully versus partial automation or rigid syntax versus natural language processing. Finally, we handled principles concerning the user's emotions, personalized experiences and security and privacy practices under the category of user experiences.

# 3. Results
## 3.1 Analysis of the websites

As is standard across websites of organizations supporting victims of domestic abuse, 'quick exit' buttons were displayed at the top of the screen. These buttons enable survivors to exit the website quickly with one mouse click in an emergency case. They are presented in bright colors (usually red) and move up and down on the webpage when scrolling to remain visibility in the same place. All websites stated that the first step that should be taken in an emergency is contacting the police.

### 3.1.1 Communication Techniques

Organizations follow similar approaches to interact with survivors. Phone numbers and email accounts were the most common addresses that survivors are directed to. Sending offline messages was another option offered to the survivors where they could fill forms to explain the problem and the way that the organization could contact them safely. Other personal identifiers such as names were not given as mandatory fields to fill. Communication via WhatsApp was also supported by some organizations. A live chat option, which enabled survivors to talk to people in support teams was another service that some of the organizations provided. Generally, live chat services were limited to working hours since they require human staffing. Survivors were assured that the conversations with members of support services (especially charities) would remain confidential and would not be shared with third parties without their permission. Some of the organizations went further and guaranteed that conversations would be deleted when interaction ended.

As another communication strategy, we have identified two chatbots solutions; one for online abuse on National Domestic Abuse Helpline[28] and another for domestic abuse on Southall Black Sisters[34]. Both were button-based chatbots that provide necessary information in their own domains. Offering the user to choose from several options, which were presented in the form of menus or buttons, the chatbots displayed information depending on what the user clicked on.

### 3.1.2 Website Analysis

Analyzing website content, there were two notable categories; main messages targeted to the readers and essential information given by experts/staff. Three themes were identified among the messages showing the importance of supporting survivors emotionally; "Domestic abuse is a crime/ It is never OK", "Domestic abuse is never the fault of the person who is experiencing it." and "You're not alone". Example sites include Refuge[27] and WomensAid[40]. These awareness-raising messages were perceived as a key component of support for those impacted by domestic abuse.

We identified seven main themes for essential information provided for domestic abuse. "Risk assessment" questions are covered by almost all organizations and people in immediate need or in danger are directed to call the police. Information regarding "Types of abuse" is another significant theme. Psychological abuse, financial abuse, sexual abuse, coercive control, physical abuse, and online abuse are the most frequently explained abuse types on the websites. Readers are also provided with direction that can enable them to recognize the abuse. "Spotting signs" that allow individuals to understand if they are in an abusive relationship are available on most of the websites. This guidance is enriched with the information on "Impact of domestic abuse" mainly on health and well-being.

"Support with legal options" is another emerging theme which covers several topics including family law, criminal law, and immigration and asylum law. Information that can be summarized as "Practical advice" is another informative theme identified in the context of the websites. This category contains a variety of practical guidance on housing, making safety plans and monetary issues. Finally, we observed that some organizations dedicated to preventing domestic abuse recognized online abuse as well and provided "Guidance on online safety". Under this theme awareness of spyware is emphasized and guidance is given on private browsing, deleting cookies and address histories and finally password management.

To sum up, organizations use their websites for informing survivors about how to contact with them, sharing contents that help them to recognize abuse and providing guidance they need including legal options and making safety plans.

### *3.2 Interviews*

In this section, we will define the central themes emerging from the interview data which relate to the following subsections.

#### *3.2.1 Essential information that should be given to survivors*

Our results led to the key theme "essential information," which includes a variety of topics including where to find local services, questions around family courts, financial or emotional support and housing options, or how to leave a partner safely. In addition, it is stated that survivors also need guidance on recognizing abuse. They need validation and often the input of support organizations to assess whether a particular behavior is normal or abusive. Interviewees agree on the importance of the safety planning, risk assessment and recognition of abusive behaviors and reported that as essential information that each survivor should know. One of the participants explicitly stated that, "recognizing that you're in an abusive relationship is a primary thing". However, it is added by another participant that "perpetrators tend to undermine the abuse that's happening and not paint the real picture, but it's also a trauma defense from victim's as well, so they will minimize what's happening to them in order to be able to cope every day".

Consequently, it is crucial to provide guidance to survivors on recognition of abuse and risk assessment. The police officers interviewed pointed to two pertinent pieces of information that are used to assess the risk faced; the level of harm and frequency of it. These points are also considered within the context of if the event was currently happening or in the past. After a risk assessment, the next step mentioned is safety planning, which covers both the victims and any others affected (e.g., children).

In addition to that practical advice, we noted that some key messages were reported as being important when engaging with those impacted by domestic abuse. These messages cover the following points: "It is not your fault", "We believe you", "There is help available". This highlights the supportive nature of the interaction.

#### *3.2.2 Particularly vulnerable groups to domestic abuse*

Identification of the "vulnerable groups" to domestic abuse is important in the design of the customized chatbots that can consider needs of different individuals. Our participants emphasized the fact that domestic abuse can happen to anyone, regardless of ethnicity, religion, sexuality, background, or gender. However, they also pointed to some specific groups with unique needs that are often overlooked. Disabled women were mentioned by the all the participants as the most vulnerable group, where women with both physical and psychological disabilities have been referred. Disadvantages or difficulties in access to support have been reported as the key factor that led to increased vulnerability. Not speaking English as a first language, having a minority ethnic origin, having no recourse to public funds due to immigration status or having children (which presents challenges in finding appropriate housing) were given as other examples of relevant factors. Participants also highlighted that elderly woman (abused by adult children) as another vulnerable groups that have been often overlooked.

*3.2.3 Support that bystanders need*
In exploring the support needed by those that have witnessed domestic abuse, we noted a similar general reply as that of the responses to victims themselves. As one participant stated, "By living in a house where there is domestic abuse, they are not just witnessing it, they are experiencing it". Another added, "… that's definitely something that is lacking in terms of services. And that's largely down to resources; the fact that we're already underfunded so they can often be forgotten". Guidance on domestic abuse and nuances, and how to be supportive without putting the survivor at risk were also listed as important information that should be conveyed to such bystanders.

*3.2.4 Burdens frequently expressed by survivors during reporting their cases to police*
Three burdens were mentioned as the most frequently expressed by survivors during reporting their cases to police; fear of not being taken seriously or believed, fear of retribution and feeling left out of the system. More importantly, participants stated that most of the cases were not reported due to several barriers including fear of causing more trouble, difficulty in providing evidence, self-blame and threats towards children or other family numbers. Difficulties in recognizing abuse and minimizing what has happened were the other barriers survivors experience impacting upon whether they report or not.

*3.2.5 How chatbots can support survivors of domestic abuse*
Considering the needs of survivors, difficulties they experience and the variety of the vulnerable groups, we asked our participants to explain how they envisaged chatbots supporting survivors of domestic abuse. The most common suggestion emerging was that such agents may be best placed as providing practical advice or factual information regarding domestic abuse and key support systems available (including those best suited and closer to individuals). Legal remedies (guidance on where to go for legal help or how to protect children), information about local services, safety planning and Universal Credit were types of the factual information that fit this use case. Participants stated that this could significantly lower the manual effort of human support teams while still providing an informative, engaging experience.

Given the importance of the recognition of domestic abuse, participants agreed on the use of chatbots for clarifying the signs for abuse. Pointing people in the appropriate direction and encouraging them to report to the police or to seek help from national help lines were also highlighted by interviewees. Police officers especially emphasized the reluctance of survivors to come to the police in the first place due to the series of formal activities/events, and they agreed with the other participants on the ability of chatbots to support and encourage individuals. They noted that a suitable role for chatbots may be as a 24/7 preface to the police/national help lines/charities; such an interface would allow escalation as appropriate, particularly if an individual is in imminent danger. A crucial point here is that the appropriateness of this direction will highly depend on the details of the cases and may not be straightforward; this needs to be considered by the chatbot.

A few participants argued for the use of chatbots in conveying emotional support. Validating survivor's experiences and reassuring them that they are believed and supported were given as examples in this context. Assuring confidentiality and not asking personal identifiers were other noted design guidelines. A key suggestion is to talk about privacy practices at the beginning of the conversation. It was argued that this would help build trust between agent and the survivor before the interaction and encourage survivors to disclose more. None of the interviewees were open to artificial intelligence-based solutions and highlighted the potential significant consequences of failures in agents giving victims inappropriate responses. The police officers interviewed, in particular, cited various concerns in failing to adequately support victims, and demonstrated clear resistance to fully automated, AI solutions. Instead, rule based chatbots were recommended. This is a key finding in our work as it can inform future research in the area, especially as it relates to the utility of large language model (LLM) systems like ChatGPT, or Generative AI more broadly.

*3.2.6 Essential traits of chatbots*
Empathy, non-judgment, friendliness, and warmth were listed as most essential traits for chatbots to be used in support of survivors. Participants agreed that chatbots can remove the feeling of judgement that is often present in those reporting, i.e., feeling that they are judged by the ones who they are reporting to. Within this context, police officers pointed to male victims and stated that chatbots may be particularly helpful for males while reporting domestic abuse since males often may feel judged or embarrassed in talking to someone. Therefore, a chatbot may increase the likelihood of reporting as it is not a human.

Assuring confidentiality and making it clear that the person is speaking to a computer application were identified as issues that should be covered within chatbot-human conversations. Interviewees also noted that agents should empower individuals to make decisions by validating their experiences and assuring them that domestic abuse is a crime. Participants experienced in face-to-face or live chat support highlighted that chatbots should ideally have names and face images (giving different ethnicity options) to assure more warmth in the conversations. Text-based chatbots were encouraged instead of voiced-based ones due to safety reasons. Specifically, any solutions that could easily be noticed by the perpetrator should be avoided since it can put the survivor at more risk.

Finally, participants underlined the difficulties faced by those disclosing domestic abuse, and recommended questions that chatbots should ask to reassure and avoid overloading individuals. Some examples which may help to present agents in a more helpful and understanding way were: "How are you feeling at the moment", "Do you want to carry on" and "Do you need to take a break?".

*3.3 Reflecting on design guideline literature*
In this phase of the study, we identified the design principles of the chatbots informed by a systematic literature review of relevant studies. As explained in the Analysis of design guideline literature, we have categorized the principles followed in the proposed solutions or suggested in the literature, into three categories; user-chatbot interaction, chatbot development and user experience.

*3.3.1 User-chatbot Interactions*
Chatbots proposed in the literature to support survivors of domestic abuse are mainly text-based[3,12,20,25] or button-based[30,39] solutions. Use of an avatar is preferred by one of the studies[30].

*3.3.2 Chatbot Development*
AI-based solutions which can build natural dialog systems automatically were generally avoided. Semi-automated approaches were proposed in one of the solutions where the automation is limited to identification of harassment type[3]. As the type is identified, responses are generated by rule-based scripts.

*User Experience*
The most frequently cited personality trait that the chatbots were expected to have to work in this domain is empathy. It was followed by lack of judgement, sympathy, and friendliness, which were stated as crucial to comfort survivors during the conversation. Importance of language support was highlighted by one of the studies due to the difficulties that victims (e.g., potentially those with a complex immigration status) can experience while seeking help[39]. Transparency about data privacy policies, data storage and third-party sharing is further recommended by another study[23]. Moreover, barriers such as stigma, confidentiality, and fear of retaliation contribute to low rates of reporting or counseling. Hence, data privacy principles are crucial in supporting better engagement.

Consent management is another factor mentioned in the literature[3]. Survivors should be able to decide whether agents can keep their conversation records or whether it should be deleted upon exit. For the individuals who agree to save their conversation records, it is critical that the chatbot explains who owns the data and who will be held responsible in the case of any information leakage. Collecting consent to keep anonymized data for further use (e.g., training data) is also noted as an essential part of consent management. Other design practices to assure confidentiality include asking as little personally identifiable information as possible or applying anonymization techniques on personal data collected.

As it relates to the services that should be offered by the chatbots to support survivors, the most frequently cited service was targeted at motivating persons to report cases to the police. Guidance during crisis, practical information on how to report or how to collect evidence are also other important services. In one of the studies[23], it is suggested that chatbots deliver personalized information depending on time of incident, location, and assault categories. The argument here is that this would be essential to provide effective support in emergency cases where chatbots should notify survivors to immediately go to the hospital (e.g., with a message containing the locations of the nearest hospitals).

**4. Towards design principles**
This section draws on the analysis of the results to outline a series of design principles for chatbots to support victims/survivors of domestic abuse and violence. These are therefore informed by interviews with key support staff who support victims, an analysis of websites which often act as the first point of contact for those impacted, and academic literature.

*4.1 Potential use cases*
Based on our findings, the use cases that appear to be the most promising in supporting victims are providing practical information and more general guidance on domestic abuse. Analysis of the interviews highlighted that recognition of abuse is extremely important to break the pattern of domestic abuse. Hence, comprehensive guidance on different types of abuse (e.g., physical abuse, sexual abuse, emotional or psychological abuse, financial abuse) and the signs of them should be covered during the conversation to better help survivors. As suggested by participants, chatbots providing such kinds of information could effectively be used as a preface to face-to-face or live chat services with human support staff. More generally, these agents may also help to reduce demands on such staff, particularly in times such as the current coronavirus (COVID-19) pandemic.

In the light of our findings, we propose following design principles/considerations:
(1) The support provided by the chatbots should cover practical information especially for safety planning (where to find local services, questions around housing options or how to leave a partner safely).
(2) Supporting survivors with information on the legal options available to them may be an ideal use case for chatbots.
(3) Information about all types of abuse should be included in the provided guidance and survivors should be informed about the signs of each type of domestic abuse.

(4) Chatbots may be used as a preface to face-to-face or live chat services. They can guide and encourage people to seek appropriate human support.
(5) Information conveyed by chatbots should include guidance on online safety which covers information regarding awareness of spyware, private browsing, deleting cookies and address histories and password management).
(6) Chatbots may be particularly useful in cases where victims may avoid reporting because they feel that they would be judged. This relates to all victims including females, males and those that are non-binary, and those in various forms of heterosexual or same-sex relationships.
(7) Chatbots may not be ideal for a risk assessment service, due to difficulties in properly categorizing cases and the potential for a significant impact if an inaccurate assessment is conducted.

*4.2 Dialog Design*

While dialog structure is studied in the literature for contexts such as customer services[13] or citizen participation[36], there were no detailed guidelines proposed for domestic abuse. However, we were able to extract a series of abstract guidelines from the literature and based on interviewees' insights. We summarize these as follows:
(1) Survivors should be informed that they are talking to a computer application at the beginning of the conversation.
(2) Survivors should be provided access to help lines and the police at the beginning of the conversation; this particularly relates to people who are in immediate need or in danger.
(3) Survivors should be assured of the confidentiality of their data before being asked for any personal information.
(4) Building trust before the conversation will help to facilitate better engagement. In addition to providing data privacy practices, survivors should be given evidence about the help the chatbot could provide to users.
(5) Survivors should have the option of being reminded of data privacy practices at any stage.
(6) Survivors should be encouraged to report their cases to the police and reminded that they can seek human support at any point throughout the conversation.
(7) Chatbots should assure survivors that domestic abuse is a crime, and it is never okay.
(8) It should be made clear by the chatbots that domestic abuse is never the fault of the person who is experiencing it.
(9) Survivors should be informed about the organizations/people who can help them and given the message that they are not alone.
(10) Some questions should be asked intermittently to assure survivors; "How are you feeling at the moment?", "Do you want to carry on?", "Do you need to take a break?".
(11) Chatbots should provide language support considering the needs of individuals in immigrant status.
(12) Dialog generation should be rule based. AI based solutions may not be ideal due to risky consequences of failure of chatbots to give appropriate responses.

It is possible to argue that the most challenging issue around dialog design is how to manage consent and communicate data privacy issues while assuring the friendliness and warmth that are crucial for effective communication with survivors. Asking consent in formal language may fit in some use cases like finance applications well, however it may risk the acceptance of a chatbot in other use cases.[4] We believe that domestic abuse is one of those cases and designing consent practices need to be communicated in a delicate way without raising concerns of users and overwhelming them. This is another research gap identified in the literature that needs further research.

*4.3 Personality traits*

Personality traits proposed in the literature and the ones recommended by the participants during the interviews were in line with each other and lead us to formulate the following design guidelines.
(1) Empathy, non-judgment, friendliness, and warmth are essential traits for chatbots to be used in support of survivors.
(2) Chatbots should have names and face images to assure more warmth in the conversations. Different ethnicity options should be made available for face images where users can pick the ones they prefer and should be informed that they are talking to a computer application at the beginning of the conversation.

*4.4 Safety*

Our findings led us to propose following design principles for safety:
(1) Personalized chatbots are not ideal for safety reasons.
(2) Any personal information requested should be minimal and relate only to supporting appropriate responsive actions (e.g., directing to nearby hospitals or police stations).
(3) Consent management is critical[42]. Survivors should be able to decide whether the chatbots can keep the record of the conversation or any personal data collected.
(4) If stored, content of conversations should be anonymized or encrypted.
(5) Chatbots should be able to explain who owns the data collected during a conversation, and who will be held responsible in the case of any information leakage.
(6) Text-based agent services are preferred in cases of domestic abuse support. This is particularly to maintain the safety of the victim (e.g., not being heard by other members of the household when seeking support).
(7) Any content provided to those impacted by domestic abuse (e.g., legal guidance, local services, etc.) should be kept up to date.
(8) Chatbots should be launched within websites and should provide 'quick exit' options (e.g., buttons or dedicated keystrokes). They should not be designed as standalone applications that require download or storage due to risks of perpetrator access.

*4.5 Unique needs of victims that should be considered during the design of chatbots*

From our analysis of the data, vulnerable groups identified within this study whose needs should be identified and considered for effective chatbot design are also listed below in our proposed guidelines:
(1) Chatbots should consider the needs of disabled people.
(2) The elderly and young women and the ones in same sex relationships are other vulnerable groups that may be supported by a chatbot solution.
(3) Solutions should be inclusive and consider needs of male victims or persons identifying as non-gender binary.
(4) Bystanders, who witness the violence, also need support; this may be another area where chatbots can be utilized.

**5. Conclusion**

Chatbots are widely available and accessible to anyone with Internet access. Their popularly has become even more salient since the debut of ChatGPT and other LLMs. Nowadays, they are mainly utilized in personal assistance, health, and finance sectors. However, they have a potential to facilitate support of survivors of domestic abuse due to their lack of judgement or ability to assure anonymity of users. In this study, we investigated current technology used in support of survivors, conducted interviews with experts in this domain and identified design features covered in the literature for agents developed to serve survivors. Finally, based on our findings we proposed a set of initial design principles for chatbots to be used to support survivors. We believe this is an important step and one which sets the foundation for development of appropriate support agents, and for more empirical research in this domain.

We are aware that our research is mainly UK-centric and reflects the opinions of professionals working for UK-based charities and information and services provided by their organizations. Even though our literature review covers the solutions proposed in different countries, interviews conducted with professionals working in developing countries where women rights are at risk may differentiate the results due some other concerns or needs of the survivors in those countries. We are also aware that current technology use in the analyzed websites might be limited since primary focus of those websites is not the use of high-tech solutions. More sophisticated practices can be identified via focus groups that include experts from both support services and the chatbot industry.

Our contributions can be grouped in to five main categories. Firstly, we identified the potential use cases where chatbots can play a role to support survivors. We have also detected the ones that are not ideal for chatbot interaction. Secondly, we proposed dialog designs which provides chatbot creators a prescriptive knowledge with a set of design principles and conversation flow. Personality traits of a chatbot that is dedicated to work with this very specific group provides our third contribution. Considering the needs of the survivors, assuring their safety while offering help is essential and as such, we also identified design principles for safety and privacy, which will help to develop reliable solutions. Finally, we identified the vulnerable groups that are often overlooked by researchers and developers of technological solutions. We believe that our findings will enable and encourage chatbot creators to develop solutions for survivors and they can lead effective use of those agents with the aim of supporting them. This will serve our ultimate purpose; increasing the number of people who can access the help they need and deserve.

**Acknowledgments**
This work was supported by the UK Engineering and Physical Sciences Research Council (EPSRC) under the title of `A Platform for Responsive Conversational Agents to Enhance Engagement and Disclosure (PRoCEED)' (EP/S027211/1 and EP/S027297/1).

**References**
[1] Alaa A Abd-Alrazaq, Mohannad Alajlani, Ali Abdallah Alalwan, Bridgette M Bewick, Peter Gardner, and Mowafa Househ. 2019. An overview of the features of chatbots in mental health: A scoping review. International Journal of Medical Informatics 132 (2019), 103978.
[2] Citizen Advice. [n.d.]. Domestic violence and abuse - getting help. https://www.citizensadvice.org.uk/family/gender-violence/domestic-violenceand-abuse-getting-help/
[3] Tobias Bauer, Emre Devrim, Misha Glazunov, William Lopez Jaramillo, Balaganesh Mohan, and Gerasimos Spanakis. 2019. #MeTooMaastricht: Building a chatbot to assist survivors of sexual harassment. In Joint European Conference on Machine Learning and Knowledge Discovery in Databases. Springer, 503–521.
[4] Rahime Belen Sağlam and Jason R C Nurse. 2020. Is your chatbot GDPR compliant? Open issues in agent design. In Proceedings of the 2nd Conference on Conversational User

Interfaces. (2020).

[5] Virginia Braun and Victoria Clarke. 2006. Using thematic analysis in psychology. Qualitative research in psychology 3, 2 (2006), 77–101.

[6] Scottye J Cash, Lauren Murfree, and Laura Schwab-Reese. 2020. "I'm here to listen and want you to know I am a mandated reporter": Understanding how text message-based crisis counselors facilitate child maltreatment disclosures. Child abuse & neglect 102 (2020), 104414.

[7] Rape Crisis. 2021. Rape Crisis: England and Wales. https://rapecrisis.org.uk/

[8] DataQuest. 2020. Safecity develops a bot on Gupshup to help women report sexual harassment on chat. https://www.dqindia.com/safecitydevelops-a-bot-on-gupshup-to-help-women-report-sexual-harassment-on-chat/

[9] Ahmed Fadhil and Gianluca Schiavo. 2019. Designing for health chatbots. arXiv preprint arXiv:1902.09022 (2019).

[10] FLOWS. [n.d.]. FLOWS-Finding Legal Options for Women Survivors. https://www.rcjadvice.org.uk/family/flows-finding-legal-options-for-womensurvivors/

[11] Office for National Statistics. 2020. Domestic abuse in England and Wales overview: November 2020. https://www.ons.gov.uk/peoplepopulationandcommunity/crimeandjustice/bulletins/domesticabuseinenglandandwalesoverview/november2020

[12] Fernando Galdon and SJ Wang. 2019. Designing trust in highly automated virtual assistants: A taxonomy of levels of autonomy. In International Conference on Industry, Vol. 4.

[13] Ulrich Gnewuch, Stefan Morana, and Alexander Maedche. 2017. Towards Designing Cooperative and Social Conversational Agents for Customer Service.. In ICIS.

[14] HelloCass. [n.d.]. HelloCass. https://hellocass.com.au/

[15] Honour Network Helpline. 2019. Karma Nirvana - Supporting victims of honour based abuse and forced marriage. https://karmanirvana.org.uk/

[16] National LGBT+ Domestic Abuse Helpline. 2017. Galop the LGBT+ anti-violence charity. http://www.galop.org.uk/domesticabuse/

[17] National Stalking Helpline. [n.d.]. National Stalking Helpline - Suzy Lamplugh Trust. https://www.suzylamplugh.org/pages/category/nationalstalking-helpline

[18] ManKind Initiative. 2021. ManKind Initiative: Helping men escape domestic abuse. https://www.mankind.org.uk/

[19] Respect Men's Advice Line. 2021. Respect - Men's Advice Line. https://www.respect.uk.net/

[20] Adam S Miner, Arnold Milstein, Stephen Schueller, Roshini Hegde, Christina Mangurian, and Eleni Linos. 2016. Smartphone-based conversational agents and responses to questions about mental health, interpersonal violence, and physical health. JAMA internal medicine 176, 5 (2016), 619–625.

[21] Rights of Women. 2014. Rights of Women: Helping women through the. law. https://rightsofwomen.org.uk/

[22] Action on Elder Abuse. [n.d.]. Action on Elder Abuse - The National Care Line. https://www.thenationalcareline.org/AccessingHelp/ ActionOnElderAbuse

[23] Hyanghee Park and Joonhwan Lee. 2020. Can a Conversational Agent Lower Sexual Violence Victims' Burden of Self-Disclosure?. In Extended Abstracts of the 2020 CHI Conference on Human Factors in Computing Systems. 1–8.

[24] Respect Phoneline. 2019. Help For Domestic Violence Perpetrators: Respect Phoneline. https://respectphoneline.org.uk/

[25] Mihiata Pirini, Dawn Duncan, Bridgette Toy-Cronin, and David Turner. 2020. An Evaluation of Legal Information Chatbots: Useability, Utility, and Accuracy. (2020).

[26] Everyman Project. 2019. Everyman Project. http://www.everymanproject.co.uk/
[27] Refuge. 2017. Recognising Abuse. https://www.refuge.org.uk/get-help-now/recognising-abuse/
[28] Refuge. 2019. Refuge: For Women and Children against domestic violence. https://www.nationaldahelpline.org.uk/
[29] Ebony Rempel, Lorie Donelle, Jodi Hall, and Susan Rodger. 2019. Intimate partner violence: A review of online interventions. Informatics for health and social care 44, 2 (2019), 204–219.
[30] Jingjing Ren, Timothy Bickmore, Megan Hempstead, and Brian Jack. 2014. Birth control, drug abuse, or domestic violence? What health risk topics are women willing to discuss with a virtual agent?. In International Conference on Intelligent Virtual Agents. Springer, 350–359.
[31] Respond. 2019. Respond: from hurting to healing. https://respond.org.uk/
[32] RCJ Advice Family Service. [n.d.]. RCJ Advice. https://www.rcjadvice.org.uk/family/
[33] SignHealth Domestic Abuse Service. [n.d.]. SignHealth - Domestic Abuse Service. https://signhealth.org.uk/with-deaf-people/domestic-abuse/
[34] Southall Black Sisters. 2020. Southall Black Sisters. https://southallblacksisters.org.uk/
[35] SurvivorsUK. 2021. SurvivorsUK:male rape and sexual abuse. https://www.survivorsuk.org/
[36] Navid Tavanapour, Mathis Poser, and Eva AC Bittner. 2019. Supporting the Idea Generation Process in Citizen Participation-toward an Interactive System with a Conversational Agent as Facilitator.. In ECIS.
[37] UnWomen. 2019. Using AI in accessing justice for survivors of violence. https://www.unwomen.org/en/news/stories/2019/5/feature-using-ai-inaccessing-justice-for-survivors-of-violence
[38] UnWomen. 2020. Facts and figures: Ending violence against women. https://www.unwomen.org/en/what-we-do/ending-violence-againstwomen/facts-and-figures
[39] Isadora Varejao. 2019. The Women Against Violence Experiment (WAVE): social journalism solutions to help immigrant women protect themselves against abuse. (2019).
[40] WomensAid. 2020. Challenging the Myths. https://www.womensaid.org.uk/information-support/what-is-domestic-abuse/myths/
[41] WomensAid. 2020. Domestic abuse is a gendered crime. https://www.womensaid.org.uk/information-support/what-is-domestic-abuse/domesticabuse-is-a-gendered-crime/
[42] Belen-Saglam, R., Nurse, J. R. C., & Hodges, D. (2022). An investigation into the sensitivity of personal information and implications for disclosure: A UK perspective. Frontiers in Computer Science, 4, 908245.